\documentclass[10pt, conference]{IEEEtran}

\IEEEoverridecommandlockouts
\usepackage{cite}
\usepackage{amsmath,amssymb,amsfonts}
\usepackage{algorithmic}
\usepackage{graphicx}
\usepackage{textcomp}
\usepackage{xcolor}
\usepackage{ifthen}
\usepackage{booktabs}
\usepackage{nicefrac}
\usepackage[hidelinks]{hyperref}
\usepackage{scalefnt}
\usepackage{color, colortbl}
\usepackage[flushleft]{threeparttable}
\usepackage[square,comma,numbers,sort]{natbib}
\usepackage{balance} 
 
\newcommand{\MyBox}[1]{\vspace{4mm}\noindent\framebox[\columnwidth][c]{\parbox[b]{0.95\columnwidth}{ #1 }}\vspace{4mm}}

\newcommand{\MyPara}[1]{\vspace{.2em}\noindent\textit{\textbf{#1}}\hspace{.3em}}

\definecolor{Gray}{gray}{0.9}
\def\BibTeX{{\rm B\kern-.05em{\sc i\kern-.025em b}\kern-.08em
  T\kern-.1667em\lower.7ex\hbox{E}\kern-.125emX}}
 
 \AtBeginDocument{%
  \providecommand\BibTeX{{%
    \normalfont B\kern-0.5em{\scshape i\kern-0.25em b}\kern-0.8em\TeX}}}
    
\newcommand*{\img}[1]{%
    \raisebox{-.3\baselineskip}{%
        \includegraphics[
        height=\baselineskip,
        width=\baselineskip,
        keepaspectratio,
        ]{#1}%
    }%
}

\usepackage{xcolor}

\begin{document}

\title{How Do Software Developers Use GitHub Actions to Automate Their Workflows?}

%

\author{\IEEEauthorblockN{
Timothy Kinsman\IEEEauthorrefmark{1},
Mairieli Wessel\IEEEauthorrefmark{2},
Marco A. Gerosa\IEEEauthorrefmark{3} and
Christoph Treude\IEEEauthorrefmark{1}}
\IEEEauthorblockA{\IEEEauthorrefmark{1}University of Adelaide, Australia}
\IEEEauthorblockA{\IEEEauthorrefmark{2}University of São Paulo, Brazil}
\IEEEauthorblockA{\IEEEauthorrefmark{3}Northern Arizona University, USA\\
timothy.kinsman@student.adelaide.edu.au, mairieli@ime.usp.br, \\ marco.gerosa@nau.edu, christoph.treude@adelaide.edu.au}
}

\maketitle

\begin{abstract}

Automated tools are frequently used in social coding repositories to perform repetitive activities that are part of the distributed software development process. Recently, GitHub introduced GitHub Actions, a feature providing automated workflows for repository maintainers. Although several Actions have been built and used by practitioners, relatively little has been done to evaluate them. Understanding and anticipating the effects of adopting such kind of technology is important for planning and management. Our research is the first to investigate how developers use Actions and how several activity indicators change after their adoption. Our results indicate that, although only a small subset of repositories adopted GitHub Actions to date, there is a positive perception of the technology. Our findings also indicate that the adoption of GitHub Actions increases the number of monthly rejected pull requests and decreases the monthly number of commits on merged pull requests. These results are especially relevant for practitioners to understand and prevent undesirable effects on their projects.

\end{abstract}


\section{Introduction}

Social coding platforms, such as GitHub, have changed the collaborative nature of open source software development by integrating mechanisms such as issue reporting and pull requests into distributed version control tools~\cite{Dabbish2012,gousios2014exploratory}. This pull-based development workflow offers new opportunities for community engagement but at the same time increases the workload for repository maintainers to communicate, review code, deal with contributor license agreement issues, explain project guidelines, run tests, and merge pull requests~\citep{Gousios2016}.

To reduce this intensive workload, developers often rely on automation tools to perform repetitive tasks to check whether the code builds, the tests pass, and the contribution conforms to a defined style guide~\citep{kavaler2019tool}. GitHub projects adopt, for example, tools to support Continuous Integration and Continuous Delivery or Deployment (CI/CD)~\cite{zhao2017impact,cassee2020silent} and for code review~\cite{kavaler2019tool}. In recent years, software bots have been widely adopted to automate a variety of predefined tasks around pull requests~\cite{Wessel2018}. By automating part of the workflow, developers hope to increase both productivity and quality~\cite{Vasilescu2015}.

To further support automation, GitHub recently introduced GitHub Actions\footnote{https://github.com/features/actions} (the feature was made available to the public in November 2019). GitHub Actions allow the automation of tasks based on various triggers (e.g., commits, pull requests, issues, comments, etc.) and can be easily shared from one repository to another, making it easier to automate how developers build, test, and deploy software projects.

However, little is known about the impact of such kind of automation and the challenges it might impose on the project development process. In this paper, we aim to understand how software developers use GitHub Actions to automate their workflows and how the dynamics of pull requests of GitHub projects change following the adoption of GitHub Actions. To achieve our goal, we address the following research questions:

\textbf{RQ1:} \textit{How do OSS projects use GitHub Actions?}

We aim to understand how commonly repositories use GitHub Actions and what they use them for. As a results of this analysis, we found a small subset of active repositories (0.7\% of the 416,266 repositories) adopted GitHub Actions. These Actions are spread across 20 categories, including continuous integration, utilities, and deployment. We also analyzed the commit history of files related to GitHub Action workflows to understand how the use of predefined Actions evolves over time. Overall, we found that a typical Action is added two times, and never removed or modified.

\textbf{RQ2:} \textit{How is the use of GitHub Actions discussed by developers?}

To gain an insight into how developers perceive GitHub Actions, we manually analyzed a set of 209 GitHub issues that discuss GitHub Actions. We found distinct categories of discussions related to GitHub Actions' maintenance and implementation, including switching other automation tools to Actions, suggestions to implement Actions, and problems and frustrations.

\textbf{RQ3:} \textit{What is the impact of GitHub Actions?}

In this RQ, we investigate whether project activity indicators, such as the number of pull requests merged and non-merged, number of comments, the time to close pull requests, and number of commits change after GitHub Actions adoption. We used a \textit{Regression Discontinuity Design}~\cite{thistlethwaite1960regression} to model the effect of Action adoption across 926 projects that had adopted GitHub Actions for at least 6 months. Our findings indicate that, on average, there are more rejected pull requests and fewer commits on merged pull requests after adopting GitHub Actions.

In summary, we make the following contributions: (i) bringing attention to GitHub Actions, a relevant yet neglected resource that offers support for developers' tasks; (ii) characterizing the usage of GitHub Actions, and (iii) providing an understanding of how GitHub Actions' adoption impacts project activities and what developers discuss about them. 

\section{Workflow Automation with GitHub Actions}

\begin{figure*}[!htbp]
\scriptsize
\centering
\includegraphics[scale=0.7]{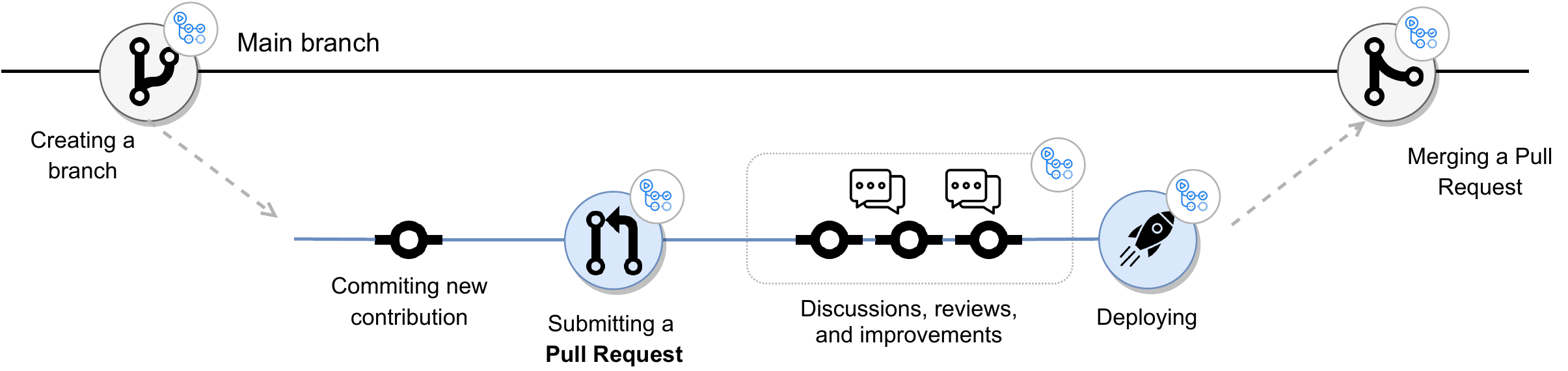}
\caption{GitHub workflow automation with GitHub Actions (adapted from GitHub).}
\label{fig:flow}
\end{figure*}

GitHub Actions is an event-driven API provided by the GitHub platform to automate development workflows. GitHub Actions can run a series of commands after a specified event has occurred. An event is an specific activity that triggers a workflow run, as shown in Figure~\ref{fig:flow} (see the \img{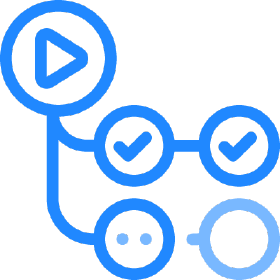} icon). For example, a workflow is triggered when a pull request is created for a repository or when a pull request is merged into the main branch. Workflows are defined in the \textbf{.github/workflows/} directory and use YAML syntax, having either a .yml or .yaml file extension. 

A workflow can contain one or more Actions. Developers can create their own Actions by writing custom code that interacts with their repository, and use them in their workflows or publish them on the GitHub Marketplace. GitHub allows developers to build Docker and JavaScript Actions and both require a metadata file to define the inputs, outputs, and main entry point of the Action.

After the successful execution of a workflow, the outputs can be displayed in different ways. One of the possibilities is through a \textit{GitHub Action bot}. This bot, as any other bot on GitHub, is implemented as a GitHub user that can submit code contributions, interact through comments, and merge or close pull requests~\cite{wessel2020inconvenient}. Recently, developers published GitHub Action variants for many well-known bots (e.g., Coveralls, Codecov, Snyk) and these Actions are rapidly increasing in popularity~\cite{golzadeh2020groundtruth}.

\begin{figure}[!htbp]
\scriptsize
\centering
\includegraphics[scale=0.6]{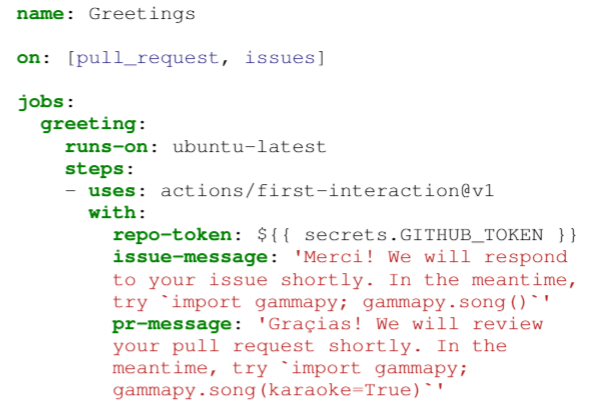}
\caption{Greetings workflow of Gammapy -- \textit{greetings.yml}}
\label{listing:greetings}
\end{figure}

\begin{figure}[!htbp]
\scriptsize
\centering
\includegraphics[scale=0.55]{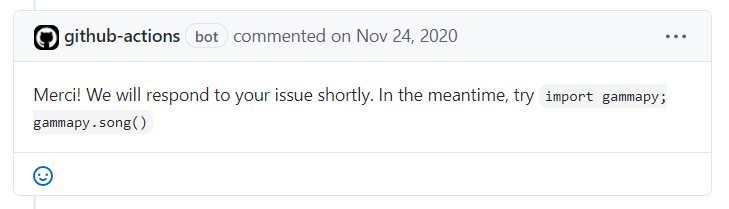}
\caption{Example of \textit{github-actions} bot greeting a newcomer.}
\label{fig:greetings-example}
\end{figure}

As an example of GitHub Actions adoption, consider the case of the project \textit{Grammapy}\footnote{https://github.com/gammapy/gammapy}, an open-source Python package for gamma-ray astronomy. As of the 13$^{th}$ of November, 2019, the \textit{Grammapy} community adopted a GitHub Action called \textit{First Interaction}\footnote{https://github.com/marketplace/actions/first-interaction}, which is responsible for identifying and welcoming newcomers when they create their first issue or open their first pull request on a project. As shown in Listing~\ref{listing:greetings}, \textit{Grammapy} created a workflow called \textit{Greeting} that might be triggered by both new pull requests and issues, as defined by the \textbf{on} keyword. The output of the \textit{First Interaction} Action is displayed through an issue/pull request comment posted by \textit{GitHub Action Bot} when a new pull request or issue is authored by a new contributor. An example of this Action interaction on a GitHub issue is shown in Figure~\ref{fig:greetings-example}.

\section{Research Design}

This study aims to understand GitHub Actions usage in GitHub projects. In the following, we present our study design, data collection, and analysis procedures.

\subsection{Selecting Projects}

We assembled a dataset of GitHub open-source projects that adopted GitHub Actions at some point in their history. To compose our study sample, we started by selecting repositories from GitHub. For this, we used the GitHub project metadata of Munaiah et al.'s \cite{munaiah2017curating} RepoReapers data set, which contained 446,862 GitHub repositories classified as containing an engineered software project.

We then filter this dataset to keep open-source software projects that at some point had adopted a GitHub Action. To identify these projects, we retrieved data from the GitHub API using a Ruby toolkit called Octokit.rb.\footnote{http://octokit.github.io/octokit.rb/}  We verified whether the repositories had files of yaml format in the \textit{./github/workflows} directory. This filtered dataset comprises 3,190 projects.

\subsection{Analyzing the use of GitHub Actions}

First, we collected and quantitatively analyzed the number of projects using GitHub Actions and the number of Actions per project (\textbf{RQ1}). We also analyzed the workflow files of the studied projects searching for the category, description, and whether the Action was verified by GitHub. To understand the evolution of GitHub Actions, we retrieve the commit history of each workflow file used by the studied projects. For this purpose, we compared the commit history looking for changes regarding Actions, which include addition, removal, configuration change, or version update. 

\subsection{Categorizing GitHub Actions Discussions}

To answer \textbf{RQ2}, we gathered issues from the repositories that mention either ``github action'' or ``github actions'', have at least one comment, and were posted after the release of the GitHub Actions feature. We collected 209 issues that met these criteria. After collecting these issues, we manually analyzed and categorized them. Two researchers independently conducted the manual classification. The first author of this paper conducted the manual classification of the 209 issues. Another researcher categorized a subset of 25 random issues using the same model. We scored a free-marginal kappa value of 0.66. We then conducted a second negotiation round and scored a free-marginal kappa value of 0.76. Fleiss et al.~\cite{fleiss2013statistical} state that the rule of thumb is that kappa values less than 0.40 are poor, values from 0.40 to 0.75 are intermediate to good, and values above 0.75 are excellent. 

\subsection{Time series analysis}

To answer \textbf{RQ3}, we conducted a time series analysis. We collected longitudinal data for different outcome variables and treated the adoption of GitHub Actions by each project in our data set as an ``intervention''. This way, we could align all the time series of project-level outcome variables on the intervention date and compare their trends before and after adopting Actions. 

In the following subsections, we detail the different steps involved, from filtering the initial data set to running the statistical models.

\subsubsection{Aggregating project variables}

We analyzed data from 6 months before and 6 months after the Action adoption. Similarly to previous work~\cite{zhao2017impact,wessel2020effects,cassee2020silent}, we exclude 30 days around the Action adoption date to avoid the influence of the instability caused during this period. Afterward, we aggregated individual pull request data into monthly periods, considering 6 months before and after the Action introduction. Afterward, we checked the activity level of the candidate projects, since many projects on GitHub are inactive~\cite{gousios2014exploratory}. We removed from our dataset (i) projects that did not received any pull requests or (ii) that disabled the Actions during the period we considered. After applying all filters, our data set comprises 926 active projects that have been using at least one GitHub Action for 6 months. We focused on the same pull request related variables as in previous work \cite{wessel2020effects}:

\MyPara{Merged/non-merged pull requests:} the number of monthly contributions (pull requests) that have been merged, or closed but not merged into the project, computed over all closed pull requests in each time frame.

\MyPara{Comments on merged/non-merged pull requests:} the median number of monthly comments computed over all merged and non-merged pull requests in each time frame.

\MyPara{Time-to-merge/time-to-close pull requests:} the median of monthly pull request latency (in hours), computed as the difference between the time when the pull request was closed and the time when it was opened. The median is computed using all merged and non-merged pull requests in each time frame.

\MyPara{Commits of merged/non-merged pull requests:} the median of monthly commits computed over all merged and non-merged pull requests in each time frame.

Based on previous work~\cite{cassee2020silent,zhao2017impact,wessel2020effects}, we also collected six known covariates for each project:

\MyPara{Project name:} the name of the project to which the pull request belongs. This name is used to uniquely identify the project on GitHub.

\MyPara{Programming language:} the primary project programming language, as automatically provided by GitHub.

\MyPara{Time since the first pull request:} in months, computed since the earliest recorded pull request in the entire project history. We use this variable to capture the project maturity when it comes to the pull request usage.

\MyPara{Total number of pull request authors:} we count how many contributors submitted pull requests to the project as a proxy for the size of the project community.

\MyPara{Total number of commits:} we compute the total number of commits as a proxy for the activity level of a project.

\MyPara{Number of pull requests opened:} the number of monthly contributions (pull requests) received in each time frame. We expect that projects with a high number of contributions also observe a high number of comments, latency, commits, and merged and non-merged contributions.

\subsubsection{Statistical Approach}
\label{sec-statistical-modeling}

We modeled the effect of GitHub Action adoption over time across GitHub repositories using a Regression Discontinuity Design (RDD)~\cite{thistlethwaite1960regression,imbens2008regression}, following the work of Wessel et al. \cite{wessel2020effects}. RDD is a technique used to model the extent of a discontinuity at the moment of intervention and long after the intervention. The technique is based on the assumption that if the intervention does not affect the outcome, there would be no discontinuity, and the outcome would be continuous over time~\cite{quasiexperimentation}. 
The statistical model behind RDD is
\begin{equation*}
\begin{split}
y_{i} =&\: \alpha + \beta\cdot \mbox{\textit{time}}_{i} + \gamma\cdot \mbox{\textit{intervention}}_{i} \: + \\& \delta\cdot \mbox{\textit{time\_after\_intervention}}_{i} \: + \eta\cdot \mbox{controls}_{i} + \varepsilon_{i}
\end{split}
\end{equation*}
where $i$ indicates the observations for a given project.

To model the passage of time as well as the GitHub Action introduction, we rely on three variables: \textit{time}, \textit{time after intervention}, and \textit{intervention}. The \textit{time} variable is measured as months at the time $j$ from the start to the end of our observation period for each project. We considered a time period of 12 months for this study, 6 months before and after bot adoption.

The \textit{intervention} variable is a binary value used to indicate whether the time $j$ occurs before ($\mbox{\textit{intervention}}=0$) or after the ($\mbox{\textit{intervention}}=1$) adoption event. The \textit{time\_after\_intervention} variable counts the number of months at time $j$ since the Action adoption, and the variable is set to 0 before adoption. The $\mbox{\textit{controls}}_{i}$ variables enable the analysis of Action adoption effects, rather than confounding the effects that influence the dependent variables. For observations before the intervention, holding controls constant, the resulting regression line has a slope of $\beta$, and after the intervention $\beta+\delta$. The size of the intervention effect is measured as the difference equal to $\gamma$ between the two regression values of $y_{i}$ at the moment of the intervention. 

Considering that we are interested in the effects of GitHub Actions on the monthly trend of the number of pull requests, number of comments, time-to-close pull requests, and number of commits for both merged and non-merged pull requests, we fitted eight models ($4$ variables $\times$ $2$ cases). To balance false-positives and false-negatives, we report the corrected p-values after applying multiple corrections using the method of Benjamini and Hochberg~\cite{benjamini1995controlling}. We implemented the RDD models as a mixed-effects linear regression using the R package \textit{lmerTest}~\cite{kuznetsova2017lmertest}.

Following the work of Wessel et al.~\cite{wessel2020effects}, we modeled \textit{project name} and \textit{programming language} as random effects~\cite{galecki2013linear} to capture project-to-project and language-to-language variability~\cite{zhao2017impact}. We evaluate the model fit using \textit{marginal} $(R^2_m)$ and \textit{conditional} $(R^2_c)$ scores, as described by Nakagawa and Schielzeth~\cite{nakagawa2013general}. The $R^2_m$ can be interpreted as the variance explained by the fixed effects alone, and $R^2_c$ as the variance explained by the fixed and random effects together.

In mixed-effects regression, the variables used to model the intervention along with the other fixed effects are aggregated across all projects, resulting in coefficients useful for interpretation. The interpretation of these regression coefficients supports the discussion of the intervention and its effects, if any. Thus, we report the significant coefficients ($p < 0.05$) in the regression as well as their variance, obtained using ANOVA. In addition, we \textit{log} transform the fixed effects and dependent variables that have high variance~\cite{sheather2009modern}. We also account for multicollinearity, excluding any fixed effects for which the variance inflation factor (VIF) is higher than $5$~\cite{sheather2009modern}.

\section{Results}

In the following, we report the results of our study per research question.

\subsection{How do OSS projects use GitHub Actions (RQ1)?}
\label{howmany}

\begin{figure}[!htbp]
\scriptsize
\centering
\includegraphics[scale=0.45]{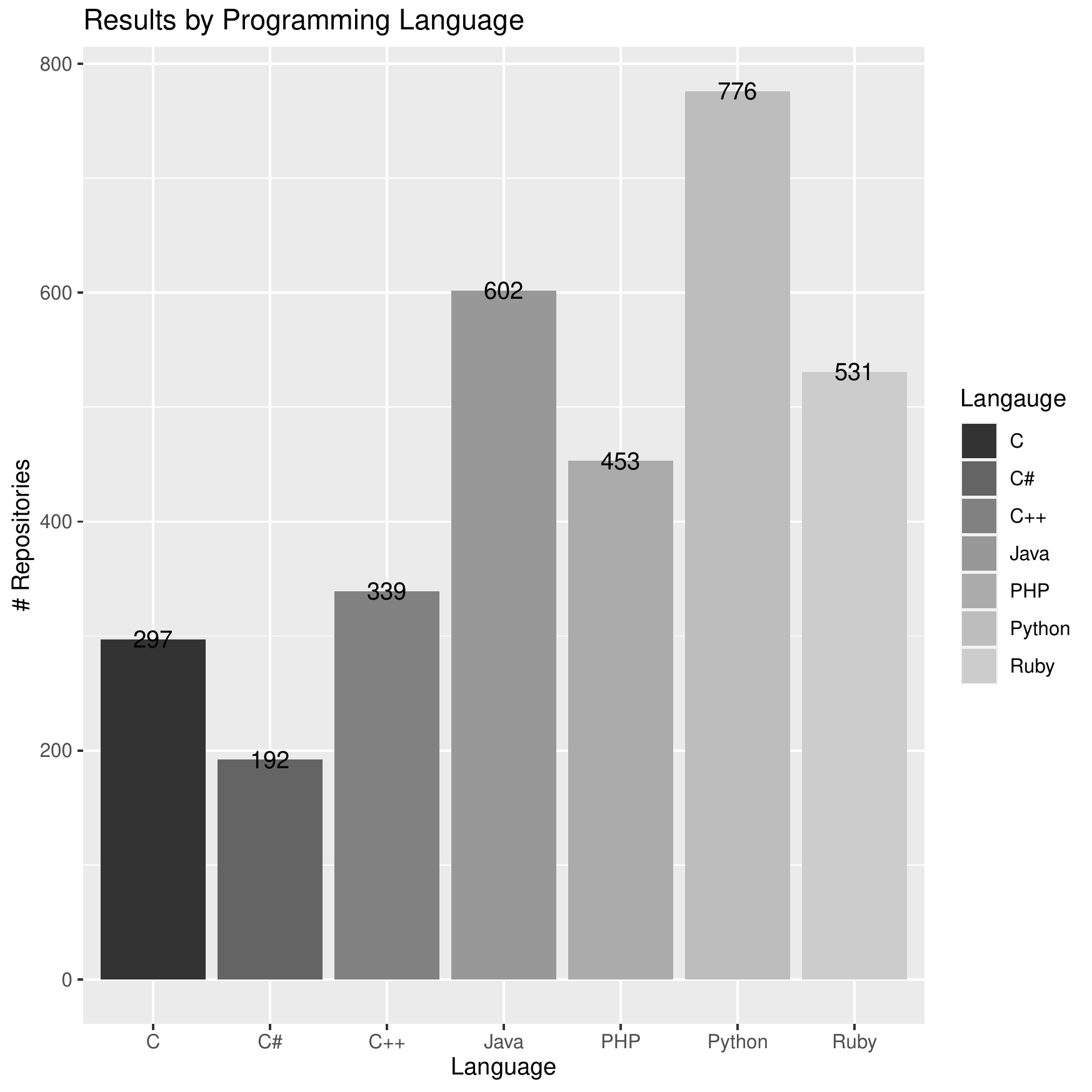}
\caption[Number of repositories that use GitHub Actions]{Number of repositories that use GitHub Actions.}
\label{fig:Bar}
\end{figure}

Analyzing a set of 416,266 active (i.e., received pull requests in the relevant time frame, see previous section) repositories, we identified 3,190 (0.7\%) open-source software projects that had adopted at least one GitHub Action at the time of our data collection. Figure~\ref{fig:Bar} reports the absolute number of repositories that use GitHub Actions grouped by programming language, showing that the most prominent adopters of GitHub Actions are Python repositories, followed by Java and Ruby.

We collected the data only 10 months after GitHub Actions was released to the public and our data show that a number of projects had already adopted the technology.
Of the 3,190 GitHub repositories that use GitHub Actions, we found a total of 708 different predefined Actions. 

We collected data from each Action's repository and also from the GitHub Marketplace\footnote{https://github.com/marketplace?type=actions} page to categorize the Actions. If published in the marketplace, an Action is classified in 1--2 categories by the publisher. Table~\ref{tab:categories} presents the categorization of Actions we found. Note that the percentages do not add up to 100 since about half of the Actions are assigned to two categories, a primary one and a secondary one.

\begin{table}[!htbp]
\centering
\caption[Categorization of Actions found within GitHub Actions workflows]{Categorization of Actions found within GitHub Actions workflows.}
\begin{tabular}{lrr}
\hline 
\textbf{Actions' Categories}    & \textbf{\# of Actions} & \textbf{\%}  \\ \hline
Continuous integration & 192              & 27.12        \\
Utilities              & 173              & 24.44        \\
Deployment             & 87               & 12.29        \\
Publishing             & 70               & 9.89         \\
Code quality           & 53               & 7.49         \\
Code review            & 45               & 6.36         \\
Dependency management  & 36               & 5.08         \\
Testing                & 33               & 4.66         \\
Open Source management & 30               & 4.24         \\
Project management     & 27               & 3.81         \\
Container CI           & 25               & 3.53         \\
Chat                   & 18               & 2.54         \\
Security               & 13               & 1.84         \\
Community              & 6                & 0.85         \\
Desktop tools          & 5                & 0.70         \\
Mobile                 & 5                & 0.70         \\
Mobile CI              & 4                & 0.56         \\
IDEs                   & 3                & 0.42         \\
Monitoring             & 3                & 0.42         \\
Localization           & 2                & 0.28         \\
Uncategorised          & 280              & 39.55        \\ \hline
\textbf{total Actions} & \textbf{708}     & \textbf{156.77} \\ \hline
\end{tabular}
\label{tab:categories}
\end{table}

The five most frequent categories of Actions are the following:

\MyPara{Continuous integration:} Actions responsible for running the CI pipeline and notifying contributors of test failures in CI tools  (e.g., Retry Step, Chef Delivery).

\MyPara{Utilities:} Actions created to automate diverse steps of the development workflow on the GitHub platform, often in support of other Actions. The \textit{Read Properties} Action, for example, inspects Java \textit{.properties} files looking for predefined properties. Another example of a utility Action is \textit{Replace string}, which replaces strings that match predefined regular expressions.

\MyPara{Deployment:} Actions designed to build and deploy the application upon request. One example is the Action called \textit{Jekyll Deploy}, responsible for building and deploying the Jekyll site to GitHub Pages.

\MyPara{Publishing:} Actions responsible for automatically publishing packages to the registry. For example, \textit{Action For Semantic Release} is an Action that leverages \textit{semantic-release} to fully automate the package release workflow, determining the next version number, generating the release notes and publishing the package.

\MyPara{Code quality:} Actions that analyze source code (e.g., code style, code coverage, code quality, and smells) submitted through pull requests and give feedback to developers via GitHub checks or comments.

In addition, we found that 42 (5.93\%) out of 708 Actions are verified by GitHub. Creators are verified if they have an existing relationship with GitHub, and GitHub has worked closely with the creator to create these Actions.

The five most popular Actions are the following:

\MyPara{actions/checkout:} A verified, utility Action that checks-out a repository under \$GITHUB\_WORKSPACE. Therefore, a workflow can access the repository for further workflow tasks.

\MyPara{actions/setup-python:} A verified, utility Action that sets up a Python environment for use in a workflow, allowing the use of Python features and commands.

\MyPara{actions/cache:} A verified utility and dependency management Action that allows caching dependencies and building outputs to improve workflow execution time.

\MyPara{actions/upload-artifact:} A verified utility Action that uploads artifacts from a workflow, allowing developers to share data between jobs and store data once a workflow is complete.         

\MyPara{actions/setup-java:} A verified utility action that sets up a Java environment for use in a workflow, allowing the use of Java features and commands, such as compiling and executing.



Analyzing the version histories of these GitHub repositories, we found that in addition to adding a GitHub Action, repositories also removed Actions, modified their arguments, and updated their versions. We investigated how often these events occur and which Actions were most affected.

\begin{table}[t]
\caption{GitHub Actions that were added most often.}
\label{tab:additions}
\centering
\begin{tabular}{lrl}
\toprule
Action & \# & Description \\
\midrule
actions/checkout          & 7,962 & Check out a repository  \\
actions/setup-python      & 1,756 & Set up workflow with Python  \\
actions/cache             & 1,729 & Cache dependencies/build outputs  \\
actions/upload-artifact   & 1,441 & Upload artifacts from workflow  \\
actions/setup-java        & 877  & Set up workflow with Java  \\
actions/download-artifact & 580  & Download artifacts from build  \\
shivammathur/setup-php    & 434  & Set up workflow with PHP  \\
actions/setup-ruby        & 373  & Set up workflow with Ruby  \\
codecov/codecov-action    & 253  & Upload coverage to Codecov \\
actions/setup-dotnet      & 225  & Set up workflow with .NET \\
\bottomrule
\end{tabular}
\end{table}

Table~\ref{tab:additions} shows the top 10 GitHub Actions that were added to a repository most often. At the median, Actions were added two times (average: 30).

\begin{table}[t]
\caption{GitHub Actions that were removed often.}
\label{tab:removals}
\centering
\begin{tabular}{lrl}
\toprule
Action & $\nicefrac{rm}{add}$ & Description \\
\midrule
masa-iwasaki/setup-rbenv           & 1.00 & Rbenv setup \\
meeDamian/github-release           & 0.94 & Github Releases \\
eregon/use-ruby-action             & 0.81 & Prebuilt Ruby \\
jakejarvis/s3-sync-action          & 0.77 & Sync with S3 \\
harmon758/postgresql-action        & 0.73 & PostgreSQL setup \\
kiegroup/github-action-build-chain & 0.67 & Build multiple projects \\
alexjurkiewicz/setup-ccache        & 0.67 & Ccache setup \\
actions/labeler                    & 0.64 & Pull request labelling \\
SamKirkland/FTP-Deploy-Action      & 0.64 & FTP server deploy \\
coverallsapp/github-action         & 0.62 & Coveralls upload  \\
\bottomrule
\end{tabular}
\end{table}

Naturally, the Actions that were added most often are also the ones that were removed most often. Instead of absolute numbers, we therefore analyzed the relative frequency of removals, i.e., how often different Actions were removed compared to how often they were added. Table~\ref{tab:removals} shows the ten Actions that were removed most often in relative terms. We limited this and the following analyses in this section to Actions that were added at least ten times. As the table shows, \texttt{masa-iwasaki/setup-rbenv} was removed in all cases where it had previously been added. Looking at the Action's README file,\footnote{\url{https://github.com/masa-iwasaki/setup-rbenv}} this is unsurprising since it contains the note `Do Not Use This Action'. Other Actions were removed less frequently, with a median of zero removals (average: 5). Some of the Actions in the top 10 have several issues reported against them regarding not being able to run on some operating systems.\footnote{e.g., \url{https://github.com/meeDamian/github-release/issues/20}}

\begin{table}[t]
\caption{GitHub Actions with many arguments modified.}
\label{tab:arguments}
\centering
\begin{tabular}{l@{}rl}
\toprule
Action & $\nicefrac{mod}{add}$ & Description \\
\midrule
andstor/copycat-action                            & 5.45 & File copying \\
reactivecircus/android-emulator-runner            & 0.62 & Android Emulators \\
julianoes/Publish-Docker-Github-Action            & 0.51 & Publish docker \\
archive/github-actions-slack                      & 0.50 & Messages to Slack \\
nanasess/setup-chromedriver                       & 0.45 & ChromeDriver setup \\
elgohr/Publish-Docker-Github-Action               & 0.43 & Publish docker \\
google/oss-fuzz                                   & 0.42 & Fuzz testing \\
docker/build-push-action                          & 0.41  & Docker with Buildx \\
actions/setup-dotnet                              & 0.38 & .NET setup \\
SamKirkland/FTP-Deploy-Action                     & 0.36 & FTP server deploy \\               
\bottomrule
\end{tabular}
\end{table}

Table~\ref{tab:arguments} shows the GitHub Actions which had their arguments modified most often, in relative terms. For example, the arguments of \texttt{andstor/copycat-action}, an Action to copy files from a repository to another external repository, were changed 5.45 times as often as the Action was added. This Action has 15 arguments, including ones to indicate the source and destination of the copy. At the median, GitHub Actions had their arguments modified zero times, with an average of 3.

\begin{table}[t]
\caption{GitHub Actions with many versions changed.}
\label{tab:versions}
\centering
\begin{tabular}{l@{}rl}
\toprule
Action & $\nicefrac{vc}{add}$ & Description \\
\midrule
gaurav-nelson/github-action & & \\
\hspace{1em}-markdown-link-check & 1.79 & Link checker \\
SamKirkland/FTP-Deploy-Action                   & 0.91 & FTP server deploy \\
technote-space/get-diff-action                  & 0.75        & Git diff \\
actions/github-script                           & 0.71 & GitHub via JavaScript  \\
homoluctus/slatify                              & 0.64 & Slack Notifications \\
stefanzweifel/git-auto-commit-action            & 0.60         & Automatically Commit \\
crazy-max/ghaction-docker-buildx                & 0.60         & Docker with Buildx \\
peter-evans/create-pull-request                 & 0.53 & Pull request creation   \\
cirrus-actions/rebase                           & 0.50         & Rebase pull requests \\
puppetlabs/action-litmus\_parallel              & 0.47 & Workflow files org. \\
\bottomrule
\end{tabular}
\end{table}

Some GitHub Actions were also frequently updated (average: 3, median: 0). Table~\ref{tab:versions} shows the Actions that had their versions modified most often, compared to how often they were added. The Action at the top of this list is under active development, with nine releases in the past six months at the time of writing.

\MyBox{\textbf{Answer to RQ1.} We identified 3,190 active GitHub repositories which have adopted the GitHub Actions feature. We found 708 unique predefined Actions being used within the workflows. These Actions are spread across 20 categories. The most recurrent ones are continuous integration, utilities, and deployment. A typical (median) GitHub Action is added twice, and never removed or modified. Some of the Actions are removed, their arguments modified, and their versions changed many times, which might be explained by their characteristics, such as release history or number of arguments.}

\subsection{How is the use of GitHub Actions discussed by developers? (RQ2)}

We categorized 209 GitHub issues based on the content of the discussion. Table~\ref{tab:issues} shows an overview of this categorization, indicating how many issues we found in each category. We present the categories in the following.

\begin{table}[!htbp]
\centering
\caption[Categorization of discussion]{Categorization of discussion.}
\begin{tabular}{lrr}
\hline
\textbf{Issues' Categories} & \multicolumn{1}{l}{\textbf{\# of Issues}} & \multicolumn{1}{l}{\textbf{\%}} \\ \hline
GitHub Actions maintenance                  & 43                                  & 20.57                           \\
Announcement GitHub Actions  & 35                                  & 16.74                           \\
Requesting GitHub Actions to be implemented & 34                                  & 16.27                           \\
Switching CI/CD tools to GitHub Actions & 32                                  & 15.31                           \\
GitHub Actions problems and frustrations      & 31                                  & 14.83                           \\
Other                                       & 31                                  & 14.83                           \\ \hline
\textbf{sum}                                & \textbf{209}                        & \textbf{100}                    \\ \hline
\end{tabular}
\label{tab:issues}
\end{table}

\MyPara{GitHub Actions maintenance:} The most recurrent topic discussed by open-source contributors and maintainers regarding GitHub Actions is their maintenance. Within this category, we classified issues which developers used to discuss about maintaining GitHub Actions, adding or requesting features to pre-existing GitHub Actions, quick fixes, workarounds, and requesting admin rights. One maintainer, for example, opened an issue to request changes in an Action responsible for reporting diffs: ``\textit{At the moment, GitHub Action for diff report generation is triggered by PR changes and comment addition. [...] Pull request should be removed and only comment trigger is left.}'' Another example relates to issues pointing to small fixes in the README file, for example, as a result of moving to GitHub Actions. 

\MyPara{Announcement of GitHub Actions:} Another recurrent topic discussed by open-source developers on GitHub issues is when a new Action is announced. This category also comprises issues announcing that GitHub Actions has been implemented (not replacing pre-existing CI/CD platforms).

\MyPara{Requesting GitHub Actions to be implemented:} We also found issues suggesting GitHub Actions to be looked into or requesting the use of GitHub Actions. We found, for example, a developer requesting to create an Action to add a label ``\textit{had PR}'' to an issue once the pull request that solves a specific issue is submitted.

\MyPara{Switching CI/CD tools to GitHub Actions:} Developers often create issues to discuss switching from a pre-existing CI/CD platform (e.g., CircleCI, Jenkins, TravisCI) to GitHub Actions (not including implementing GitHub Actions in parallel with pre-existing CI/CD tools).

\MyPara{GitHub Actions problems and frustrations:} This category encompasses bugs, broken builds, errors, and frustrations related to GitHub Actions. There are bugs caused by a failure of a service the Action relies on. For example, a maintainer opened an issue to report that the ``\textit{GitHub Actions on Mac and Windows fail due to missing Numpy}.''

\MyPara{Other:} Issues within this category relate to bugs pointed out by GitHub Actions, noise, or other discussions that do not fall into other categories.

\MyBox{\textbf{Answer to RQ2.} Overall, discussions involving problems and frustrations are outweighed by announcements that GitHub Actions had been implemented, requesting implementation, and switching CI/CD tools.}

\subsection{What is the Impact of GitHub Actions? (RQ3)} 

To answer this question, we investigated the effects of GitHub Action adoption on project activities along four dimensions: (i) merged and non-merged pull requests, (ii) human conversation, (iii) efficiency to close pull requests, and (iv) modification effort. We start by investigating how Action adoption impacts the number of merged and non-merged pull requests. We fit two mixed-effect RDD models, as described in Section \ref{sec-statistical-modeling}. For these models, the \textit{number of merged/non-merged pull requests} per month is the dependent variable. Table~\ref{tab:resultspullrequest} summarizes the results of these models. In addition to the model coefficients, the table also shows the sum of squares, with a variance explained for each variable. 

\begin{table}[htbp]
\scalefont{0.9}
\centering
\caption{The Effects of GitHub Actions on PRs. The response is \textbf{log(number of merged/non-merged PRs)} per month.}
\label{tab:resultspullrequest}
\begin{threeparttable}
\begin{tabular}{lrrrrrr}
\midrule
 & \multicolumn{2}{c}{Merged PRs} & & \multicolumn{2}{c}{Non-merged PRs}\\
\cmidrule{2-3}\cmidrule{5-6}
 & Coeffs & Sum Sq. & & Coeffs & Sum Sq. \\
\cmidrule{2-3}\cmidrule{5-6}
Intercept &  -0.203***  & & & -0.159** &\\
TimeSinceFirstPR &  0.0002 &  47.7 & & -0.001** & 6.95  \\
log(TotalPRAuthors) & -0.002 &  638.6 & & 0.028*** & 133.69 \\
log(TotalCommits) & 0.020*** &  236.5 & & 0.017** & 34.65 \\
log(OpenedPRs) & 0.770*** & 3393.8 & & 0.230*** & 530.86 \\
log(PRComments) &  0.048*** & 48.9 & & 0.342*** & 410.05 \\
log(PRCommits) & 0.246*** & 105.8 & & 0.200*** & 73.24 \\
time & 0.004 & 1.0 & & -0.004* &  0.01 \\
interventionTrue & 0.014 & 0.1 & & 0.002 & 0.03 \\
time\_after\_intervention & -0.004 & 0.2 && 0.008** & 0.47 \\
\midrule
Marginal $R^2$ & 0.88 &  &  & 0.64 & \\
Conditional $R^2$ &  0.91 &  &  & 0.78 & \\
\midrule
\end{tabular}
\begin{tablenotes}
 \item *** $p < 0.001$, ** $p < 0.01$, * $p < 0.05$
\end{tablenotes}
\end{threeparttable}
\end{table}

Analyzing the model for merged pull requests, we found that the fixed-effects part fits the data well ($R^2_m=0.88$). However, considering $R^2_c=0.91$, variability also appears from project-to-project and language-to-language. Among the fixed effects, we note that the number of monthly pull requests explains most of the variability in the model, indicating that projects receiving more contributions tend to have more merged pull requests, with other variables held constant. None of the Action-related predictors have statistically significant effects, meaning the trend in the number of merged pull requests is stationary over time, and remains unaffected by the Action adoption.

Similarly to the previous model, the fixed-effect part of the non-merged pull requests model fits the data well ($R^2_m=0.64$), even though a considerable amount of variability is explained by random effects ($R^2_c=0.78$). We  note  similar results on fixed effects: projects receiving more contributions tend to  have  more  non-merged  pull  requests. In addition, pull requests receiving more comments tend to be rejected. The effect of Action adoption on the non-merged pull requests differs from the previous model. Regarding the time series predictors, the model did not detect any discontinuity at adoption time. However, the negative trend in the number of non-merged pull requests before the Action adoption is reversed, toward an increase after adoption.

\begin{table}[htbp]
\scalefont{0.9}
\centering
\caption{The Effects of GitHub Actions on Pull Request Comments. The response is \textbf{log(median of comments)} per month.}
\label{tab:resultscomments}
\begin{threeparttable}
\begin{tabular}{lrrrrrr}
\midrule
 & \multicolumn{2}{c}{Merged PRs} & & \multicolumn{2}{c}{Non-merged PRs}\\
\cmidrule{2-3}\cmidrule{5-6}
 & Coeffs & Sum Sq. & & Coeffs & Sum Sq. \\
\cmidrule{2-3}\cmidrule{5-6}
Intercept &  0.020 & & & -0.058** &\\
TimeSinceFirstPR & -0.001*** & 3.29 & & -0.001*** & 10.98 \\
log(TotalPRAuthors) &  0.055*** & 47.49 & & 0.031*** & 169.33 \\
log(TotalCommits) & -0.008 & 2.71 & & 0.002 & 25.67 \\
log(OpenedPRs) & -0.005 & 40.87 &  & 0.051*** & 217.05 \\
log(TimeToClosePRs) &  0.077*** & 376.95 & &  0.113*** & 930.91 \\
log(PRCommits) & 0.213*** & 70.48 & & 0.199*** & 75.38 \\
time & -0.009*** & 1.54 & & 0.002 & 0.00 \\
interventionTrue & 0.001 & 0.02 & & -0.002 & 0.01 \\
time\_after\_intervention &  0.009*** &  0.70 && -0.003 & 0.05 \\
\midrule
Marginal $R^2$ & 0.38 & & &  0.65 & \\
Conditional $R^2$ & 0.54 & & & 0.70 & \\
\midrule
\end{tabular}
\begin{tablenotes}
 \item *** $p < 0.001$, ** $p < 0.01$, * $p < 0.05$
\end{tablenotes}
\end{threeparttable}
\end{table}

To investigate the effects of Action adoption on pull request communication, we fit one model to merged pull requests and another to non-merged ones. The \textit{median of pull request comments} per month is the dependent variable. Table~\ref{tab:resultscomments} shows the results of the fitted models. Considering the model of comments on merged pull requests, we found that the combined fixed-and-random effects ($R^2_c=0.54$) fit the data better than the fixed effects ($R^2_m=0.38$), showing that most of the explained variability in the data is associated with project-to-project and language-to-language variability, rather than the fixed effects. Additionally, we also observe that time-to-close pull requests explains the largest amount of variability in the model, indicating that the communication during the pull request review is strongly associated with the time to merge it. Regarding the Action effects, we note a decreasing time baseline trend before adoption; no statistically significant discontinuity at the adoption time; and an apparent neutralization of the aforementioned time trend after adoption, as $\beta(\mbox{\textit{time}}) + \delta(\mbox{\textit{time\_after\_intervention}}) \simeq 0$.

Turning to the model of comments on non-merged pull requests, the model fits the data well ($R^2_m=0.65$) and there is also variability explained by the random variables ($R^2_c=0.70$). This model also suggests that communication during the pull request review is strongly associated with the time to reject the pull request. Table~\ref{tab:resultscomments} shows that none of the Action-related predictors have statistically significant effects, meaning the comments trend in non-merged pull requests is stationary over time, and remains unaffected by the Action adoption.

\begin{table}[htbp]
\scalefont{0.9}
\centering
\caption{The Effects of GitHub Actions on Time-to-close PRs. The response is \textbf{log(median of time-to-close PRs)} per month.}
\label{tab:resultstime}
\begin{threeparttable}
\begin{tabular}{lrrrrrr}
\midrule
 & \multicolumn{2}{c}{Merged PRs} & & \multicolumn{2}{c}{Non-merged PRs}\\
\cmidrule{2-3}\cmidrule{5-6}
 & Coeffs & Sum Sq. & & Coeffs & Sum Sq. \\
\cmidrule{2-3}\cmidrule{5-6}
Intercept &  -0.374** & & &  0.053 &\\
TimeSinceFirstPR & -0.003** & 130.3 & & 0.0004 & 322.6  \\
log(TotalPRAuthors) &  0.218*** & 1386.1 & & 0.067** & 3733.9 \\
log(TotalCommits) &  0.021 & 155.6 & & -0.028 & 530.3 \\
log(OpenedPRs) & -0.139*** & 1046.3 & & 0.089*** & 4588.4 \\
log(PRComments) &  1.528*** & 8543.4 & & 2.816*** & 23589.5 \\
log(PRCommits) & 1.520*** & 4145.8 & & 1.011*** & 1967.1 \\
time & 0.013 & 2.5 & & -0.005 &  0.3 \\
interventionTrue & -0.053 & 2.1 & &  -0.003 & 0.0 \\
time\_after\_intervention & -0.004 & 0.1 &&  0.006 &  0.3 \\
\midrule
Marginal $R^2$ & 0.47 &  & & 0.63 & \\
Conditional $R^2$ &  0.57 &  & & 0.67 & \\
\midrule
\end{tabular}
\begin{tablenotes}
 \item *** $p < 0.001$, ** $p < 0.01$, * $p < 0.05$
\end{tablenotes}
\end{threeparttable}
\end{table}

We fitted two RDD models where \textit{median of time to close pull requests} per month is the dependent variable. The results are shown in Table~\ref{tab:resultstime}. Analyzing the results to the effect of GitHub Actions on the latency to merge pull requests, we found that combined fixed-and-random effects fit the data better than the fixed effects. Although several variables affect the trends of pull request latency, communication during the pull requests is responsible for most of the variability in the data. This indicates the expected results: the more effort contributors expend discussing the contribution, the more time the contribution takes to merge. The number of commits also explains the amount of data variability, since a project with many changes needs more time to review and merge them. However, none of the Action-related predictors have statistically significant effects on the time spent to merge pull request.

Turning to the model of non-merged pull requests, we note that it fits the data well ($R^2_m=0.63$), and there is also a variability explained by the random variables ($R^2_c=0.67$). As above, communication during the pull requests is responsible for most of the variability encountered in the results. Similar to the previous model, none of the Action-related predictors have statistically significant effects on the time spent to reject pull request. 

\begin{table}[htbp]
\scalefont{0.9}
\centering
\caption{The Effects of GitHub Actions on Pull Request Commits. The response is \textbf{log(median of commits)} per month.}
\label{tab:resultscommits}
\begin{threeparttable}
\begin{tabular}{lrrrrrr}
\midrule
 & \multicolumn{2}{c}{Merged PRs} & & \multicolumn{2}{c}{Non-merged PRs}\\
\cmidrule{2-3}\cmidrule{5-6}
 & Coeffs & Sum Sq. & & Coeffs & Sum Sq. \\
\cmidrule{2-3}\cmidrule{5-6}
Intercept &  -0.020*** & && -0.075** &\\ 
TimeSinceFirstPR & 0.001** & 7.02 & & -0.00004 & 14.14 \\
log(TotalPRAuthors) & -0.060*** & 83.99 & & -0.012* & 233.35 \\
log(TotalCommits) & 0.019** & 41.86 & & 0.019*** & 71.96 \\
log(OpenedPRs) & 0.247*** & 440.55 && 0.117*** & 352.79 \\
log(PRComments) & 0.541*** & 441.40 && 0.611*** & 758.36 \\
time & 0.010*** & 1.97 && 0.002 & 0.00 \\
interventionTrue & 0.039** & 0.56 & & -0.009 & 0.07 \\
time\_after\_intervention &  -0.015*** & 3.18 && -0.002 & 0.03 \\
\midrule
Marginal $R^2$ & 0.47 &  & & 0.49 & \\
Conditional $R^2$ & 0.60 &  & & 0.54 & \\
\midrule
\end{tabular}
\begin{tablenotes}
 \item *** $p < 0.001$, ** $p < 0.01$, * $p < 0.05$
\end{tablenotes}
\end{threeparttable}
\end{table}

Finally, we studied whether Action adoption affects the number of commits made before and during the pull request review. Again, we fitted two models for merged and non-merged pull requests, where the \textit{median of pull request commits} per month is the dependent variable. The results are shown in Table~\ref{tab:resultscommits}. Analyzing the model of commits on merged pull requests, we found that the combined fixed-and-random effects ($R^2_c=0.60$) fit the data better than the fixed effects ($R^2_m=0.47$). The statistical significance of all Action-related coefficients indicates that the adoption of Actions affected the number of commits. We note an increasing trend before adoption and a statistically significant discontinuity at the adoption time. Further, the positive trend in the number of merged pull requests before the Action adoption is reversed, toward a decrease after adoption. Additionally, we can also observe that the number of pull request comments and the number of contributions per month explains most of the variability in the result. This result suggests that the more comments and pull requests there are, the more commits there will be. 

Investigating the results of the non-merged pull request model, we also found that the combined fixed-and-random effects fit the data better than the fixed effects. Similar to the previous model, the number of pull request comments per month explains most of the variability in the result. Regarding the time series predictors, the model did not detect any discontinuity at adoption time. However, the positive trend in the median of commits before the bot adoption is reversed, toward a decrease after adoption.

\MyBox{\textbf{Answer to RQ3.} After adopting GitHub Actions, on average, there are more rejected pull requests and fewer commits on merged pull requests.}

\section{Discussion}

Recently, an easy, reusable, and portable way to automate developers' workflows on GitHub was made possible by the advent of GitHub Actions. So far, the literature presents scarce evidence on the use of GitHub Actions by GitHub repositories~\cite{golzadeh2020groundtruth}. In this work, we contribute by introducing and systematizing evidence on the use, evolution, and impacts of such Actions. Our findings contribute new knowledge about how software developers use GitHub Actions. Firstly, we showed that 3,190 (0.7\%) of the 416,266 repositories in our data set had already adopted GitHub Actions at the time of analysis. In addition, we found that 708 unique predefined Actions have been used within the repositories using GitHub Actions. While 39.55\% of these Actions were uncategorized, 5.93\% were verified, indicating that the majority of the Actions on GitHub are created by the community. Uncategorized Actions are Actions that have not been published to the GitHub Marketplace, thus this percentage indicates an active community surrounding GitHub Actions.

Analyzing the historical evolution of Actions, we found that some of the Actions require maintenance after being added to a repository. While a typical (median) Action is added twice (to the same or different repositories) and never removed or modified, a significant minority of Actions are removed, have their arguments modified, or are updated to new versions. For repository maintainers, this means that adding a GitHub Action effectively adds yet another dependency to a project that needs to be maintained, might become outdated, and needs to be adjusted over time. On the other hand, this is not unexpected for a new feature, and in fact all repositories that have adopted GitHub Actions to date can be seen as early adopters. A typical Action also does not appear to require ongoing maintenance, at least not in the time window considered in our study.

Announcements that GitHub Actions had been implemented, requesting implementation and switching CI/CD tools outweighed the discussions involving problems and frustrations, thus indicating a positive perception of GitHub Actions. With a new feature such as GitHub Actions, it would not be surprising if some negative sentiment would be found in corresponding discussion forums, in particular asking why yet another feature is needed. However, anecdotally, developers seem to appreciate the premise of GitHub Actions to help standardize the use of bots on GitHub, and in fact a good portion of the discussions was about switching CI/CD tools that a repository already used to their GitHub Actions equivalent.

We found that two activity indicators have a statistically significant effect on the pull request process after the adoption of an GitHub Action. According to the regression results, the median number of rejected pull requests increases after the adoption of Actions. This may indicate that project maintainers started to have faster and clearer feedback on the pull request, helping them to identify major issues on a vast number of contributions. Moreover, GitHub Actions produce different effects on non-merged pull requests when compared to the effects of adopting software bots---Wessel et al.~\cite{wessel2020effects} reported that the introduction of bots on pull requests' reviews leads to fewer rejected pull requests.

From the regression results, we also noticed an increase in the median number of commits on merged pull requests just after bot adoption. It makes sense from the contributors' side, since the Action introduces a secondary evaluation step to the pull request. Especially at the beginning of the adoption, the Action might increase the number of commits due to the need to meet all requirements and obtain a stable code. After that, however, a decrease occurs in the median number of commits on merged pull requests per month. 

Our work has implications for researchers and practitioners. For researchers interested in software bots, it is important to understand the role of GitHub Actions in the bot landscape. Backed by GitHub, GitHub Actions are likely here to stay, and we already see evidence of existing software tools, such as test coverage tools, being integrated into and packaged as GitHub Actions. It is important to understand how such Actions affect the interplay of developers in their effort to develop software, and our study provides the first step in this direction. Additional effort is also necessary to investigate the impact for newcomers, who already face a variety of barriers~\cite{balali2018newcomers,steinmacher2015social}. Educators may also see an opportunity in GitHub Actions to build automation tools to better support their OSS assignments~\cite{pinto2017training}.

Practitioners need to make informed decisions whether to adopt GitHub Actions (or software bots in general) into their projects and how to use them effectively. Also, GitHub Actions might allow them to automate repetitive tasks in their projects with their own custom GitHub Action. Already at its current early-adopter stage, GitHub Actions provides hundreds of different Actions, potentially making it difficult for practitioners to decide which Action to use, if any. Our work provides first empirical data on which Actions are currently used, how they evolve, and what their impact can be on development processes. We hope that this work will inspire more repositories to adopt GitHub Actions for their projects.  

\section{Limitations and Threats to Validity}

In this section, we discuss the limitations and threats to validity and how we have mitigated them. For replication purposes, we made our data and source code publicly available.\footnote{https://zenodo.org/record/4626256}

\MyPara{External Validity:} Since we selected engineered software projects, our findings might not be generalized to other or all GitHub projects. One way to overcome this threat is by studying non-engineered projects hosted on GitHub. Additionally, even though we considered a large number of projects and our results indicate general trends, we recommend running segmented analyses when applying our results to a given project. In addition, we focused on the same pull request related variables as in previous work \cite{wessel2020effects,cassee2020silent,zhao2017impact}, leaving other effects and artifacts for future work.

\MyPara{Construct Validity:}
As stated by Kalliamvakou et al.~\cite{Kalliamvakou2014}, many merged pull requests appear non-merged. Since we consider the number of merged pull requests, our results may be affected by this threat. Our study can be replicated when automated ways of detecting this issue are developed. 

\MyPara{Internal Validity:}
To reduce internal threats, we applied multiple data filtering steps to the statistical models. We varied the data filtering criteria to confirm the robustness of our models. For example, we filtered projects that did not receive pull requests in all months and observed similar phenomena. We also carried out a series of placebo tests~\cite{imbens2008regression} using the same model with the adoption artificially set to different dates to confirm the model robustness. The assumption of exogeneity of the treatment might be a threat. We added several controls that might influence the independent variables to reduce confounding factors.

\section{Related Work}

There has been no study investigating GitHub Actions. However previous work has investigated other automation tools such as software bots, continuous integration, and continuous delivery.

\subsection{Software Bots}
On GitHub, software bots are often integrated into the pull request workflow \cite{Erlenhov2019}, to perform a variety of tasks. These include repairing bugs \cite{Monperrus2019}, refactoring the source code \cite{Wyrich2019}, recommending tools to help developers \cite{Brown2019} and updating outdated dependencies \cite{Mirhosseini2017}. Software bots have been proposed to support technical and social aspects of software development activities \cite{Lin2016}, such as communication and decision-making \cite{Storey2016}. Van Tonder and Le Goues~\cite{Tonder2019} believe software bots are a promising addition to a developer's toolkit as they bridge the gap between human software development and automated processes.

However, understanding how software bots’ interaction affects human developers is a major challenge. Storey et al.~\cite{Storey2016} highlight that software bots’ potential negative impact is still neglected, as the way that these software bots interact on pull requests can be disruptive and perceived as unwelcoming \cite{10.1145/3387940.3391504}. Wessel et al.~\cite{Wessel2018} investigated the usage and impact of software bots to support contributors and maintainers with pull requests. After identifying bots on popular GitHub repositories, the authors classified these bots into $13$ categories according to the tasks they perform. The third most frequently used bots are code review bots. Wessel et al.~\cite{wessel2020effects} also employed a regression discontinuity design on OSS projects, revealing that the bot adoption increases the number of monthly merged pull requests, decreases monthly non-merged pull requests, and decreases communication among developers.

\subsection{Continuous Integration and Continuous Delivery (CI/CD)}
Improving software quality and reducing risks is the main goal of CI as stated by Duvall et al. \cite{duvall2007continuous}. By introducing continuous integration to the pull request process, the findings from Vasilescu et al. \cite{Vasilescu2015} clearly point to the benefits of CI. More pull requests got processed, more were being accepted and merged and more were also being rejected. In the context of Computer Science education, rising enrollments make it difficult for instructors and teaching assistants to give adequate feedback on each student's work. Hu et al.~\cite{Hu2019} had set up a static code analyzer and a continuous integration service on GitHub to help students check code style and functionality. By implementing three bots, results found by the authors showed that more than 70\% of students think the advice given by the bots is useful and can provide significantly more feedback (six times more on average) than teaching staff. A survey by Chen et al.~\cite{940726} reports that of the hundreds of billions of dollars spent on developer wages, up to 25\% accounts for fixing bugs~\cite{940726}. Continuous integration thus holds huge potential to further reduce human effort and costs by automatically fixing bugs.

Prior work has also investigated the impact of CI and code review tools on GitHub projects~\cite{zhao2017impact,kavaler2019tool,cassee2020silent} across time. While Zhao et al.~\cite{zhao2017impact} and
Cassee et al.~\cite{cassee2020silent} 
focused on the impact of the Travis CI tool's introduction on development practices, Kavaler et al.~\cite{kavaler2019tool} turned to the impact of linters, dependency managers, and coverage reporter tools. Our work extends this literature by providing a more in-depth investigation of the effects of GitHub Actions adoption.

\section{Conclusion}

In this paper, we investigate how software developers use GitHub Actions to automate their workflows, how they discuss these Actions on the issue tracker, and what are the effects of the adoption of such Actions on pull requests. While several Actions have been proposed and adopted by the open-source software community, relatively little has been done to evaluate the state of practice. To understand the impact on practice, we collected and analyzed data from 3,190 active GitHub repositories. Further, to understand the impact on practice, we statistically analyzed a  sample of 926 open-source projects hosted on GitHub.

Firstly, the findings showed that only a small subset of 3,190 repositories used GitHub Actions. We also found that 708 unique predefined Actions were being used within the workflows. Further, we collected and analyzed GitHub Actions related issues and found that the majority of the discussion was positive. These findings indicate that GitHub Actions were met with an overall positive reception among software developers. By modeling the data around the introduction of GitHub Actions, we notice different results from merged pull requests and non-merged ones. The monthly number of commits of merged pull requests decreases after the adoption of GitHub Actions and there are also more monthly rejected pull requests.

Our findings bring to light how early adopters are using, discussing, and being impacted by GitHub Actions. Learning from those early adopters can provide insights to assist the open-source community to decide whether to use GitHub Actions and how to use them effectively.
Future work includes the qualitative investigation of the effects of adopting a GitHub Actions and the expansion of our analysis for considering the effects of different types of Actions and activity indicators. 

\section*{Acknowledgments}

This work was partially supported by the Coordenação de Aperfeiçoamento de Pessoal de Nível Superior – Brasil (CAPES) – Finance Code 001, CNPq (grant 141222/2018-2), NSF grants 1815503 and 1900903, and the Australian Research Council's Discovery Early Career Researcher Award (DECRA) funding scheme (DE180100153)

\bibliographystyle{IEEEtran}
\bibliography{paper}

\begin{thebibliography}{10}
\providecommand{\url}[1]{#1}
\csname url@samestyle\endcsname
\providecommand{\newblock}{\relax}
\providecommand{\bibinfo}[2]{#2}
\providecommand{\BIBentrySTDinterwordspacing}{\spaceskip=0pt\relax}
\providecommand{\BIBentryALTinterwordstretchfactor}{4}
\providecommand{\BIBentryALTinterwordspacing}{\spaceskip=\fontdimen2\font plus
\BIBentryALTinterwordstretchfactor\fontdimen3\font minus
  \fontdimen4\font\relax}
\providecommand{\BIBforeignlanguage}[2]{{%
\expandafter\ifx\csname l@#1\endcsname\relax
\typeout{** WARNING: IEEEtran.bst: No hyphenation pattern has been}%
\typeout{** loaded for the language `#1'. Using the pattern for}%
\typeout{** the default language instead.}%
\else
\language=\csname l@#1\endcsname
\fi
#2}}
\providecommand{\BIBdecl}{\relax}
\BIBdecl

\bibitem{Dabbish2012}
L.~Dabbish, C.~Stuart, J.~Tsay, and J.~Herbsleb, ``Social coding in {GitHub}:
  Transparency and collaboration in an open software repository,'' in
  \emph{Proceedings of the ACM 2012 Conference on Computer Supported
  Cooperative Work}, ser. CSCW '12.\hskip 1em plus 0.5em minus 0.4em\relax New
  York, NY, USA: ACM, 2012, pp. 1277--1286.

\bibitem{gousios2014exploratory}
G.~Gousios, M.~Pinzger, and A.~van Deursen, ``An exploratory study of the
  pull-based software development model,'' in \emph{Proceedings of the 36th
  International Conference on Software Engineering}.\hskip 1em plus 0.5em minus
  0.4em\relax ACM, 2014, pp. 345--355.

\bibitem{Gousios2016}
G.~Gousios, M.-A. Storey, and A.~Bacchelli, ``Work practices and challenges in
  pull-based development: The contributor's perspective,'' in \emph{Proceedings
  of the 38th International Conference on Software Engineering}, ser. ICSE
  '16.\hskip 1em plus 0.5em minus 0.4em\relax New York, NY, USA: ACM, 2016, pp.
  285--296.

\bibitem{kavaler2019tool}
D.~Kavaler, A.~Trockman, B.~Vasilescu, and V.~Filkov, ``Tool choice matters:
  {JavaScript} quality assurance tools and usage outcomes in {GitHub}
  projects,'' in \emph{Proceedings of the 41st International Conference on
  Software Engineering}.\hskip 1em plus 0.5em minus 0.4em\relax IEEE Press,
  2019, pp. 476--487.

\bibitem{zhao2017impact}
Y.~Zhao, A.~Serebrenik, Y.~Zhou, V.~Filkov, and B.~Vasilescu, ``The impact of
  continuous integration on other software development practices: a large-scale
  empirical study,'' in \emph{Proceedings of the 32nd IEEE/ACM International
  Conference on Automated Software Engineering}.\hskip 1em plus 0.5em minus
  0.4em\relax IEEE Press, 2017, pp. 60--71.

\bibitem{cassee2020silent}
N.~Cassee, B.~Vasilescu, and A.~Serebrenik, ``The silent helper: the impact of
  continuous integration on code reviews,'' in \emph{27th IEEE International
  Conference on Software Analysis, Evolution and Reengineering}.\hskip 1em plus
  0.5em minus 0.4em\relax IEEE Computer Society, 2020.

\bibitem{Wessel2018}
M.~Wessel, B.~M. de~Souza, I.~Steinmacher, I.~S. Wiese, I.~Polato, A.~P.
  Chaves, and M.~A. Gerosa, ``The power of bots: Characterizing and
  understanding bots in {OSS} projects,'' \emph{Proc. ACM Hum.-Comput.
  Interact.}, vol.~2, no. CSCW, pp. 182:1--182:19, Nov. 2018.

\bibitem{Vasilescu2015}
B.~Vasilescu, Y.~Yu, H.~Wang, P.~Devanbu, and V.~Filkov, ``Quality and
  productivity outcomes relating to continuous integration in {GitHub},'' in
  \emph{Proceedings of the 2015 10th Joint Meeting on Foundations of Software
  Engineering}, ser. ESEC/FSE 2015.\hskip 1em plus 0.5em minus 0.4em\relax New
  York, NY, USA: ACM, 2015, pp. 805--816.

\bibitem{thistlethwaite1960regression}
D.~L. Thistlethwaite and D.~T. Campbell, ``Regression-discontinuity analysis:
  An alternative to the ex post facto experiment.'' \emph{Journal of
  Educational psychology}, vol.~51, no.~6, p. 309, 1960.

\bibitem{wessel2020inconvenient}
M.~Wessel and I.~Steinmacher, ``The inconvenient side of software bots on pull
  requests,'' in \emph{2nd International Workshop on Bots in Software
  Engineering}, ser. BotSE '20, 2020.

\bibitem{golzadeh2020groundtruth}
M.~Golzadeh, A.~Decan, D.~Legay, and T.~Mens, ``A ground-truth dataset and
  classification model for detecting bots in {GitHub} issue and {PR}
  comments,'' 2020.

\bibitem{munaiah2017curating}
N.~Munaiah, S.~Kroh, C.~Cabrey, and M.~Nagappan, ``Curating {GitHub} for
  engineered software projects,'' \emph{Empirical Software Engineering},
  vol.~22, no.~6, pp. 3219--3253, 2017.

\bibitem{fleiss2013statistical}
J.~Fleiss, B.~Levin, and M.~Paik, \emph{Statistical Methods for Rates and
  Proportions}, ser. Wiley Series in Probability and Statistics.\hskip 1em plus
  0.5em minus 0.4em\relax Wiley, 2013.

\bibitem{wessel2020effects}
M.~{Wessel}, A.~{Serebrenik}, I.~{Wiese}, I.~{Steinmacher}, and M.~A. {Gerosa},
  ``Effects of adopting code review bots on pull requests to {OSS} projects,''
  in \emph{2020 IEEE International Conference on Software Maintenance and
  Evolution (ICSME)}, 2020, pp. 1--11.

\bibitem{imbens2008regression}
G.~W. Imbens and T.~Lemieux, ``Regression discontinuity designs: A guide to
  practice,'' \emph{Journal of econometrics}, vol. 142, no.~2, pp. 615--635,
  2008.

\bibitem{quasiexperimentation}
T.~Cook and D.~Campbell,
  \emph{\BIBforeignlanguage{English}{Quasi-Experimentation: Design and Analysis
  Issues for Field Settings}}.\hskip 1em plus 0.5em minus 0.4em\relax Houghton
  Mifflin, 1979.

\bibitem{benjamini1995controlling}
Y.~Benjamini and Y.~Hochberg, ``Controlling the false discovery rate: a
  practical and powerful approach to multiple testing,'' \emph{Journal of the
  Royal statistical society: series B (Methodological)}, vol.~57, no.~1, pp.
  289--300, 1995.

\bibitem{kuznetsova2017lmertest}
A.~Kuznetsova, P.~B. Brockhoff, and R.~H.~B. Christensen, ``lmertest package:
  tests in linear mixed effects models,'' \emph{Journal of Statistical
  Software}, vol.~82, no.~13, 2017.

\bibitem{galecki2013linear}
A.~Ga{\l}ecki and T.~Burzykowski, \emph{Linear mixed-effects models using R: A
  step-by-step approach}.\hskip 1em plus 0.5em minus 0.4em\relax Springer
  Science \& Business Media, 2013.

\bibitem{nakagawa2013general}
S.~Nakagawa and H.~Schielzeth, ``A general and simple method for obtaining {R2}
  from generalized linear mixed-effects models,'' \emph{Methods in ecology and
  evolution}, vol.~4, no.~2, pp. 133--142, 2013.

\bibitem{sheather2009modern}
S.~Sheather, \emph{A modern approach to regression with R}.\hskip 1em plus
  0.5em minus 0.4em\relax Springer Science \& Business Media, 2009.

\bibitem{balali2018newcomers}
S.~Balali, I.~Steinmacher, U.~Annamalai, A.~Sarma, and M.~A. Gerosa,
  ``Newcomers’ barriers... is that all? an analysis of mentors’ and
  newcomers’ barriers in {OSS} projects,'' \emph{Computer Supported
  Cooperative Work (CSCW)}, vol.~27, no.~3, pp. 679--714, 2018.

\bibitem{steinmacher2015social}
I.~Steinmacher, T.~Conte, M.~A. Gerosa, and D.~Redmiles, ``Social barriers
  faced by newcomers placing their first contribution in open source software
  projects,'' in \emph{Proceedings of the 18th ACM conference on Computer
  supported cooperative work \& social computing}, 2015, pp. 1379--1392.

\bibitem{pinto2017training}
G.~H.~L. Pinto, F.~Figueira~Filho, I.~Steinmacher, and M.~A. Gerosa, ``Training
  software engineers using open-source software: the professors' perspective,''
  in \emph{2017 IEEE 30th Conference on Software Engineering Education and
  Training (CSEE\&T)}.\hskip 1em plus 0.5em minus 0.4em\relax IEEE, 2017, pp.
  117--121.

\bibitem{Kalliamvakou2014}
E.~Kalliamvakou, G.~Gousios, K.~Blincoe, L.~Singer, D.~M. German, and
  D.~Damian, ``The promises and perils of mining {GitHub},'' in
  \emph{Proceedings of the 11th Working Conference on Mining Software
  Repositories}, ser. MSR 2014.\hskip 1em plus 0.5em minus 0.4em\relax New
  York, NY, USA: ACM, 2014, pp. 92--101.

\bibitem{Erlenhov2019}
L.~Erlenhov, F.~G. de~Oliveira~Neto, R.~Scandariato, and P.~Leitner, ``Current
  and future bots in software development,'' in \emph{Proceedings of the 1st
  International Workshop on Bots in Software Engineering}, ser. BotSE
  '19.\hskip 1em plus 0.5em minus 0.4em\relax IEEE Press, 2019, p. 7–11.

\bibitem{Monperrus2019}
M.~Monperrus, ``Explainable software bot contributions: Case study of automated
  bug fixes,'' in \emph{Proceedings of the 1st International Workshop on Bots
  in Software Engineering}, ser. BotSE '19.\hskip 1em plus 0.5em minus
  0.4em\relax Piscataway, NJ, USA: IEEE Press, 2019, pp. 12--15.

\bibitem{Wyrich2019}
M.~Wyrich and J.~Bogner, ``Towards an autonomous bot for automatic source code
  refactoring,'' in \emph{Proceedings of the 1st International Workshop on Bots
  in Software Engineering}, ser. BotSE '19.\hskip 1em plus 0.5em minus
  0.4em\relax Piscataway, NJ, USA: IEEE Press, 2019, pp. 24--28.

\bibitem{Brown2019}
C.~Brown and C.~Parnin, ``Sorry to bother you: Designing bots for effective
  recommendations,'' in \emph{Proceedings of the 1st International Workshop on
  Bots in Software Engineering}, ser. BotSE '19.\hskip 1em plus 0.5em minus
  0.4em\relax IEEE Press, 2019, p. 54–58.

\bibitem{Mirhosseini2017}
S.~Mirhosseini and C.~Parnin, ``Can automated pull requests encourage software
  developers to upgrade out-of-date dependencies?'' in \emph{Proceedings of the
  32nd IEEE/ACM International Conference on Automated Software Engineering},
  ser. ASE 2017.\hskip 1em plus 0.5em minus 0.4em\relax IEEE Press, 2017, p.
  84–94.

\bibitem{Lin2016}
B.~Lin, A.~Zagalsky, M.~Storey, and A.~Serebrenik, ``Why developers are
  slacking off: Understanding how software teams use {Slack},'' in
  \emph{Proceedings of the 19th ACM Conference on Computer Supported
  Cooperative Work and Social Computing Companion}, ser. CSCW '16
  Companion.\hskip 1em plus 0.5em minus 0.4em\relax New York, NY, USA: ACM,
  2016, pp. 333--336.

\bibitem{Storey2016}
M.-A. Storey and A.~Zagalsky, ``Disrupting developer productivity one bot at a
  time,'' in \emph{Proceedings of the 2016 24th ACM SIGSOFT International
  Symposium on Foundations of Software Engineering}, ser. FSE 2016.\hskip 1em
  plus 0.5em minus 0.4em\relax New York, NY, USA: ACM, 2016, pp. 928--931.

\bibitem{Tonder2019}
R.~van Tonder and C.~L. Goues, ``Towards s/engineer/bot: Principles for program
  repair bots,'' in \emph{Proceedings of the 1st International Workshop on Bots
  in Software Engineering}, ser. BotSE '19.\hskip 1em plus 0.5em minus
  0.4em\relax IEEE Press, 2019, p. 43–47.

\bibitem{10.1145/3387940.3391504}
M.~Wessel and I.~Steinmacher, ``The inconvenient side of software bots on pull
  requests,'' in \emph{Proceedings of the IEEE/ACM 42nd International
  Conference on Software Engineering Workshops}, ser. ICSEW'20.\hskip 1em plus
  0.5em minus 0.4em\relax New York, NY, USA: Association for Computing
  Machinery, 2020, p. 51–55.

\bibitem{duvall2007continuous}
P.~Duvall, S.~Matyas, P.~Duvall, and A.~Glover, \emph{Continuous Integration:
  Improving Software Quality and Reducing Risk}, ser. A Martin Fowler signature
  book.\hskip 1em plus 0.5em minus 0.4em\relax Addison-Wesley, 2007.

\bibitem{Hu2019}
Z.~Hu and E.~Gehringer, ``Use bots to improve {GitHub} pull-request feedback,''
  in \emph{Proceedings of the 50th ACM Technical Symposium on Computer Science
  Education}, ser. SIGCSE '19.\hskip 1em plus 0.5em minus 0.4em\relax New York,
  NY, USA: Association for Computing Machinery, 2019, p. 1262–1263.

\bibitem{940726}
S.-K. Chen, W.~K. {Fuchs}, and J.-Y. Chung, ``Reversible debugging using
  program instrumentation,'' \emph{IEEE Transactions on Software Engineering},
  vol.~27, no.~8, pp. 715--727, 2001.

\end{thebibliography}

\end{document}